\documentclass[twocolumn,english,superscriptaddress,prx,floatfix,longbibliography,aps,10pt]{revtex4-1}

\usepackage[colorinlistoftodos, color=green!40, prependcaption]{todonotes}
\usepackage{comment}

\newcommand{\change}[1]{\textcolor{black}{#1}}

\usepackage[linesnumbered,ruled,vlined]{algorithm2e}
\usepackage{graphicx}
\usepackage{dcolumn}
\usepackage{bm}

\begin{document}

\preprint{APS/123-QED}

\title{Quantum-enhanced Markov Chain Monte Carlo \\ for systems larger than your Quantum Computer}

\author{Stuart Ferguson}
 \email{S.A.ferguson-3@ed.ac.uk}
\author{Petros Wallden}%
 
\affiliation{%
 Quantum Software Lab, School of Informatics, The University of Edinburgh, Edinburgh, United Kingdom
}%

\date{\today}

\begin{abstract}

Quantum computers theoretically promise computational advantage in many tasks, but it is much less clear how such advantage can be maintained when using existing and near-term hardware that has limitations in the number and quality of its qubits. 
\change{Layden et al. 
[Nature 619, 282
(2023)] proposed a promising application by introducing a Quantum-enhanced Markov Chain Monte Carlo (QeMCMC) approach to reduce the thermalization time required when sampling from hard probability distributions.}
In QeMCMC the size of the required quantum computer scales linearly with the problem, putting limitations on the sizes of systems that one can consider. In this work we introduce a framework to coarse grain the algorithm in such a way that the quantum computation can be performed using \change{considerably} smaller quantum computers and we term the method the Coarse Grained Quantum-enhanced Markov Chain Monte Carlo (CGQeMCMC). Example strategies within this framework are put to the test, with the quantum speedup persisting while using only $\sqrt{n}$ simulated qubits where $n$ is the number of qubits required in the original QeMCMC -- a quadratic reduction in resources. The coarse graining framework has the potential to be practically applicable in the near term as it requires very few qubits to approach classically intractable problem instances\change{; in this case} only 6 simulated qubits suffice to gain advantage compared to standard classical approaches when investigating the magnetization of a 36 spin system. \change{Our method can be easily combined with other classical and quantum techniques and is adaptable to various quantum hardware specifications - in particular those with limited connectivity.}

\end{abstract}

\maketitle

\section{\label{sec:intro}Introduction}

Sampling from hard distributions is a problem ubiquitous in science, technology, finance and statistics. While there are well understood families of traditional computational methods such as the Markov Chain Monte Carlo (MCMC) \cite{geyer1992, duane1987, brooks2011} that attempt to address this challenge, the advent of quantum computing offers a promising avenue for considerably more efficient solutions in randomized algorithms\change{. This includes algorithms for Monte Carlo Integration (MCI) \cite{herbert2022, akhalwaya2023}, Quantum Metropolis Sampling (QMS) \cite{temme2011, yung2012} as well as a very general quantum speedup of Monte Carlo methods \cite{montanaro2015}.}

Current Quantum Processing Units (QPUs) are labeled as `Noisy Intermediate-Scale Quantum' (NISQ) devices, meaning that although they exhibit remarkable quantum phenomena they have strict limitations in the \change{number and quality of available qubits.} In this era where full quantum error-correction is out of reach, one must explore algorithms that are not only robust to noise but \change{are} also efficient in the number of qubits \change{required}. One algorithm that addresses the former requirement is the Quantum-enhanced Markov Chain Monte Carlo (QeMCMC) of \cite{layden2023}. It introduces an elegant way to enhance the classical MCMC by ``quantizing'' one part of the classical MCMC algorithm: namely the proposal step.

\change{The QeMCMC has some promising properties alongside its noise resilience. It is expected to exceed quadratic scaling in comparison with the best classical alternative in certain contexts, making it particularly promising for early fault-tolerant quantum processors. Additionally, QeMCMC functions as a one-shot algorithm, requiring only a single execution on a QPU per step in the hybrid algorithm by avoiding the need for expectation value estimation or state tomography.} The method has also been extended, by application to Hamiltonian Monte Carlo \cite{lockwood2024} and Variational algorithms \cite{nakano2023}.

In this paper, we propose a way to significantly reduce the qubit requirement of the QeMCMC, while maintaining \change{a scaling advantage that may lead to} quantum speedup over the classical MCMC in the low temperature limit \footnote{Here we need to stress that by quantum speed up we mean that our approach outperforms the classical MCMC. It is understood, that while the size of the quantum computer used is small, and specifically smaller than what can be classically simulated, the advantage of our algorithm can be viewed as a ``quantum inspired'' approach, while the true quantum advantage can be achieved when the QPU used is larger than what can be classically simluated.}. This is done by Coarse Graining (CG) the problem instance - an Ising model lattice - in such a way that each subset ``group'' of the lattice can be evaluated in separate quantum computations, even on different QPUs in parallel. In this way, it is predicted that one can use a QPU with as few as $O(\sqrt{n})$ \change{(where n is the number of spins in the Ising model)} qubits and still achieve a \change{scaling advantage in comparison with classical techniques.} We call this method the Coarse Grained Quantum--enhanced Markov Chain Monte Carlo (CGQeMCMC).

The way in which the QeMCMC achieves a \change{scaling} advantage is by reducing the thermalization time of a Markov chain. This is the number of steps taken to reach the stationary distribution which you wish to sample from. Here we need to make a distinction between the cost of proving that thermalization is achieved fast (a task needed to prove the performance of the algorithm) and the use of our CGQeMCMC practically for some task. For the former, the primary method of analysing thermalization is by spectral gap analysis, a well known and unambiguous figure of merit. Calculating the spectral gap, however, is an entirely  different process from MCMC and is exponentially more expensive, restricting our spectral gap analysis to $10$ spins. For the latter, we use Markov chains to estimate an example observable - magnetization - and in this case we consider a system of $36$ spins to experimentally witness quick thermalization of the CGQeMCMC. In both cases, the quantum computers part of CGQeMCMC is realised in an emulation environment (using classical devices -- see Appendix \ref{app:imp} for details of the implementation).

Explicitly, our contributions are as follows:
\begin{itemize}
  \item Propose a framework for applying course graining to the QeMCMC with the Ising model used as the test case. The framework allows the problem instance to be altered to reduce the number of qubits required and, in some cases, parallelize computation over multiple QPU.
  
  \item Numerical analysis of techniques within the CG framework through absolute spectral gap of small (up to 10) spin instances. The dependence on system size of each method is analysed, with one technique (multi-sampling with improved local groups) showing favourable scaling compared to classical techniques across different temperature ranges.
  
  \item Simulation of CGQeMCMC for larger Ising model instances up to 36 spins through the calculation of average magnetization and analysis of convergence.
  
  \item Demonstration of quantum enhancement over classical methods in the above test cases while using only $\sqrt{n}$ (perfectly simulated) qubits.
  
\end{itemize}

The paper is organized as follows: In Section \ref{sec:preliminary}, we give the necessary background, namely, the classical MCMC algorithm, the spin glass Ising model, the problem of thermalization and the QeMCMC of \cite{layden2023}. 
In Section \ref{sec:fwork_fom} we propose \change{the CGQeMCMC, a general framework for coarse graining the Ising model so that smaller quantum computers can be used to tackle larger problem instances.} We also discuss \change{the} figures of merit of simple Markov chains and the difficulties in assessing thermalization. Section \ref{sec:CGQeMCMC} discusses the particular coarse grained proposal methods employed, before they are explored numerically in Section \ref{sec:numerics}. Conclusion and future directions are given in Section \ref{sec:conc}.

\section{\label{sec:preliminary}Preliminaries} 
In this section, we introduce the classical Ising model, provide an overview of MCMC and the issue of thermalization for difficult problem instances. We then introduce the QeMCMC which improves the thermalization by utilizing quantum computation.

\subsection{The Ising model}

The archetypal test-case of random sampling algorithms such as MCMC is the ``spin glass'' Ising model. \change{Analyzing these systems is crucial due to their wide-ranging applications in fields such as materials science, biology (protein folding), and computer science (artificial neural networks).} ``Glassy'' due to their magnetic disorder in analogy with the positional disorder in chemical glass, such lattices are composed of spins that can exist in either states $s_i = \{+1,-1\}$. The state $\mathbf{s}$, composed of n spins, is thus defined $\mathbf{s} = (s_1,s_2...s_n)$. Each configuration of spins $\{1,-1\}^n$ has an associated energy ($E(\mathbf{s})$), Boltzmann probability ($\mu(\mathbf{s})$), and magnetization ($m(\mathbf{s})$) defined below:

\begin{equation}
    E(\mathbf{s})=-\sum_{j>k=1}^n J_{j k} s_j s_k-\sum_{j=1}^n h_j s_j
    \label{eqn:E}
\end{equation}

\begin{equation}
    \mu(\mathbf{s})=\frac{1}{Z} e^{-E(\mathbf{s}) / T}
    \label{eqn:mu}
\end{equation}

\begin{equation}
m(\mathbf{s})=\frac{1}{n} \sum_{j=1}^n s_j
\label{eqn:m}
\end{equation}

Where $h$ and $J$ are the field and coupling respectively that define the Ising model instance, $Z$ is the partition function $Z = \sum_{\mathbf{s}} e^{-E(\mathbf{s}) / T}$ and $T$ is a temperature parameter. The Boltzmann average magnetization $\langle m\rangle_\mu  =  \sum_{\mathbf{s}} \mu(\mathbf{s}) m(\mathbf{s})$ is an example of an observable which can be hard to calculate 
due to the size of the state space. For just 100 spins, the number of possible configurations are $2^{100} \approx 10^{30}$ and requiring an algorithm to sum over these becomes intractable. 

For systems where $J$ and $h$ are unstructured and have values of varying signs, the energy landscape is particularly rugged and hard to explore. When one wishes compute the value of 
an observable from this landscape, in the low temperature limit, very few states of low energy contribute. \change{Sampling from these contributing states is particularly difficult as they can be obscured from each other by large hamming distance  \footnote{Here, Hamming distance between two states is defined as the number of positions that the two states have different spin values.} and cost barriers. Thus, when in one state transitioning to another is typically very hard.}

\subsection{Markov Chain Monte Carlo}\label{sec:MCMC}

The MCMC has a remarkable history. Named as one of the ten most influential algorithms in history, it was developed to calculate observables of a probability distribution that is too highly dimensional to be evaluated analytically or with brute force numerics \cite{Dongarra_2000}. The first component is the Markov chain - a walk on such a highly dimensional landscape - where the probability of each step only depends on the last position. The chain is normally produced by the Metropolis-Hastings algorithm which, with some probability (hence ``Monte Carlo''), rejects or accepts update steps based on the value of the probability distribution at that point \cite{metropolis1953, fishman2013}.

MCMC is able to skirt the responsibility of ever directly calculating the expensive (due to calculation of $Z$) $\mu(s)$. Instead, it performs an approximation by exploring the state space by random walk from an arbitrary starting point. Each step in the algorithm is composed of two sub-steps: proposal and acceptance. The \textit{proposal} step determines the probability of state $\mathbf{s}^\prime$ being proposed when the system is in state $\mathbf{s}$. It is represented by the matrix $Q(\mathbf{s}^\prime|\mathbf{s})$ and can be constructed in many different ways including through the use of a quantum computer as we see in Section \ref{sec:QeMCMC}. The \textit{acceptance} step is often governed by Metropolis-Hastings (MH), where the matrix $A(\mathbf{s}^\prime|\mathbf{s})$ (Equation \ref{eqn:MH}) dictates the probability that a move is accepted \cite{hastings1970, metropolis1953}. 

\begin{equation}
A\left(\mathbf{s}^{\prime} \mid \mathbf{s}\right)=\min \left(1, \frac{\mu\left(\mathbf{s}^{\prime}\right)}{\mu(\mathbf{s})} \frac{Q\left(\mathbf{s} \mid \mathbf{s}^{\prime}\right)}{Q\left(\mathbf{s}^{\prime} \mid \mathbf{s}\right)}\right)
\label{eqn:MH}
\end{equation}

Thus, the algorithm is able to explore a complicated state space by proposing a new state to move to, and moving to that state if it is of lower energy. Of course, the Markov chain can also move to states of higher energy with probability defined by Equation \ref{eqn:MH}. Pseudo code is provided in Algorithm \ref{alg:MCMC}.

\begin{algorithm}\label{alg:MCMC}
    \caption{MCMC}
    \SetAlgoNoEnd
    \SetAlgoNoLine
    \SetNlSty{}{}{}
    
    \tcp{Initialization}
    $\mathbf{s} \gets $ initial spin configuration\;

    \While{not converged}{
        \tcp{Proposal step}
        $s^{\prime} \gets$ update proposal\;

        \tcp{Classical MH accept or reject}
        $A\left(\mathbf{s}^{\prime} \mid \mathbf{s}\right)\gets\min \left(1, \frac{\mu\left(\mathbf{s}^{\prime}\right)}{\mu(\mathbf{s})} \frac{Q\left(\mathbf{s} \mid \mathbf{s}^{\prime}\right)}{Q\left(\mathbf{s}^{\prime} \mid \mathbf{s}\right)}\right)$

        \If{$A \geq$ random.uniform$(0, 1)$}{
            $\mathbf{s} \gets \mathbf{s}^{\prime}$\;
        }
    }

\end{algorithm}

The importance of the acceptance step underlies any future algorithm. One does not need to calculate $\mu(\mathbf{s})$ or $\mu(\mathbf{s}^\prime)$ at each step, merely their ratio which can be computed in $O\left(n^2\right)$ polynomial time \cite{metropolis1953}. Together, proposal and acceptance steps result in the probability matrix, $P\left(\mathbf{s}^{\prime} \mid \mathbf{s}\right)=A\left(\mathbf{s}^{\prime} \mid \mathbf{s}\right) Q\left(\mathbf{s}^{\prime} \mid \mathbf{s}\right)$, which is the probability that each step in the algorithm will result in transition from $\mathbf{s}$ to $\mathbf{s}^\prime$. 

To ensure convergence, the Markov chain must be irreducible and aperiodic meaning that any state can be (eventually) reached from any other state and that there exists no repeating loops in the chain. \change{These are true in our case as no states are forbidden assuming there is no unwanted symmetry as discussed in \cite{layden2023}. Detailed balance is another important condition that ensures the stationary distribution of our Markov Chain is the probability distribution from which we wish to sample. It requires that the following is satisfied for any transition $\mathbf{s}\rightarrow\mathbf{s}^\prime$ where $s \ne s^\prime $.}

\begin{equation}
P(\mathbf{s}^\prime | \mathbf{s}) \mu(\mathbf{s}) = P(\mathbf{s} | \mathbf{s}^\prime) \mu(\mathbf{s^\prime}) 
\label{eqn:det_bal}
\end{equation}

In this context, the probability distribution $\mu(\mathbf{s})$ is often labeled as the ``stationary'' or ``target'' distribution. When the temperature drops the exponential term in the Boltzmann distribution becomes increasingly sensitive to small differences in energy levels. Consequently, the acceptance probability for proposed transitions becomes negligible, leading to extremely slow exploration of the state space. This phenomenon impedes efficient sampling and hinders convergence, as only proposals of similar or lower energy are consistently accepted. This means the Markov-chain in a relatively low energy state lies dormant until it happens across one of only a few low energy states, which takes exponential time. The QeMCMC aims to address this issue as explored in Section \ref{sec:QeMCMC}.

\subsection{Thermalization}\label{sec:therm}

Upon initialization, the Markov-chain proceeds until thermalization is achieved, signifying convergence to the stationary distribution. Subsequently, one can then sample from the chain at intervals spaced to ensure uncorrelated samples which is called the auto-correlation time of the Markov Chain. The expectation value of observables, such as magnetization, can then be estimated from these non-correlated samples. However, the crux lies in thermalization. A \textit{local} proposal method, altering one spin at a time, tends to get trapped in local minima. Conversely, a \textit{uniform} random proposal explores the energy landscape inefficiently, being entirely non-local. In the low temperature regime, both approaches exhibit slow auto-correlation and thermalization times. It should be noted that the uniform distribution will faithfully converge in a number of steps similar to that of the size of the entire state space. In other words, given the time it would take to solve the problem by brute force, the uniform proposal can approximate the solution. Thus, the uniform proposal provides an upper bound to the possible thermalization times of MCMC.

One of the issues with thermalization regards how one can tell when a Markov chain has reached the stationary distribution, and how many steps to take as auto-correlation time. Multiple techniques have been proposed, however many are deemed to be fickle or flawed \cite{Vehtari2021, flegal2008, geyer2011}. For example, it is possible for two different chains on the same problem instance to both individually appear like they have converged - their observables appear stable. When they are compared side by side however, it is clear they they have ``converged'' to different local distributions as the observables calculated from sampling the chains are completely different \cite{moins2023}. Alternatively, two chains may agree on observables while neither seeming to be in a stationary distribution. The former is a particular problem when updates are local. 

Classical techniques such as parallel tempering or population annealing sample multiple different chains, allowing one to parallelize computation and mitigate for the issues of convergence with local proposals \cite{machta2011, wang2015, weigel2021, swendsen1986}. These techniques are all directly applicable to CGQeMCMC, meaning any gain in convergence is propagated easily to state of the art techniques.

\subsection{Quantum enhancement}\label{sec:QeMCMC}

Recently, the QeMCMC has extended the classical MCMC to a Noisy Intermediate Scale Quantum (NISQ) computing setting, with a cubic reduction in MCMC thermalization time for some problems  \cite{layden2023}. In QeMCMC, the proposal states of the Markov chain are the outcome of a quantum computation, with speedup derived from reducing the number of required update steps. The first step is to encode the current state of the Markov Chain as binary input to the computational basis state of the quantum computer - mapping the $\{1,-1\}^n$ spin states to $\{1,0\}^n$ computational basis states. One then evolves this state according to a Hamiltonian ($H$) . 

\begin{equation}
H = (1- \gamma)\alpha H_{prob} +\gamma H_{mix}
\label{eqn:H}
\end{equation}

This unitary evolution takes the form: $U=e^{-i H t}$ with practical implementation requiring an approximation through the Trotter-Suzuki product. The problem Hamiltonian ($H_{prob}$) encodes the classical model instance and although the mixing Hamiltonian could take many different forms, it is defined here as $H_{\mathrm{mix}}=\sum_{j=1}^n X_j$ \footnote{The notation is kept in line with previous literature on the subject where possible \cite{layden2023,nakano2023}}. Here, hyper-parameters are sampled in the range $t = (2, 20)$ and $\gamma = (0.25, 0.6)$, although optimization of these parameters is an ongoing issue.

\begin{equation}
H_{\text {prob }} = -\sum_{j>k=1}^n J_{j k} Z_j Z_k -\sum_{j=1}^n h_j Z_j =\sum_{\mathbf{s}} E(\mathbf{s}) |\mathbf{s}\rangle\langle \mathbf{s}|
\label{eqn:H_prob}
\end{equation}

Measurement in the computational basis state returns a binary outcome that is used as the proposal state as seen in Algorithm \ref{alg:QeMCMC}. The Hamiltonian is inspired by the adiabatic program, and perturbative analysis expects it to propose states similar in energy to the previous state, and far in Hamming distance \cite{layden2023}. Both of these properties are optimal for fast MCMC convergence rate, and cannot be replicated using classical computers. Other choices of unitary evolution have been proposed to replace the time-independent Hamiltonian evolution, however these have not yet been explored experimentally and thus wont be considered here \cite{mazzola2021}.

\begin{algorithm}\label{alg:QeMCMC}
    \caption{QeMCMC proposal step}
    \SetAlgoNoEnd
    \SetAlgoNoLine
    \SetNlSty{}{}{}
    
    \tcp{Initialization}
    $\mathbf{s} \gets $ initial spin configuration\;
    
    \tcp{Quantum proposal step}
    $\gamma \gets$ hyper-parameter in range $(0.25, 0.6)$ \;  
    $t \gets$ hyper-parameter in range $(2, 20)$\;
    \tcp{quantum computation}
    $|\psi\rangle \gets \exp[-i H t]|\mathbf{s}\rangle$ \;
    $\mathbf{s}^{\prime} \gets$ X measurement result of $|\psi\rangle$ \;
    
\end{algorithm}

The quantum proposal distribution, $Q\left(\mathbf{s}^{\prime} \mid \mathbf{s}\right)=\left|\left\langle\mathbf{s}^{\prime}|U| \mathbf{s}\right\rangle\right|$, has the symmetry requirement: $\left|\left\langle\mathbf{s}|U| \mathbf{s}^{\prime}\right\rangle\right| = \left|\left\langle\mathbf{s}^{\prime}|U| \mathbf{s}\right\rangle\right|$. Normally, one is required to calculate $Q\left(\mathbf{s}^{\prime} \mid \mathbf{s}\right)$ explicitly as in Equation \ref{eqn:MH}, however as it is merely the ratio of $\frac{Q\left(\mathbf{s}^{\prime} \mid \mathbf{s}\right)}{{Q\left(\mathbf{s} \mid \mathbf{s}^{\prime}\right)}}$ required, the equation easily simplifies thanks to the symmetry constraint.

\begin{equation}
Q\left(\mathbf{s}^{\prime} \mid \mathbf{s}\right)=\left|\left\langle\mathbf{s}^{\prime}|U| \mathbf{s}\right\rangle\right|^2=\left|\left\langle\mathbf{s}|U| \mathbf{s}^{\prime}\right\rangle\right|^2=Q\left(\mathbf{s} \mid \mathbf{s}^{\prime}\right)
\label{eqn:symmetry}
\end{equation}

This reduces the Metropolis method (Equation \ref{eqn:MH}) to:

\begin{equation}
A\left(\mathbf{s}^{\prime} \mid \mathbf{s}\right)=\min \left(1, \frac{\mu\left(\mathbf{s}^{\prime}\right)}{\mu(\mathbf{s})} \right) = \min \left(1, \exp{\frac{E(\mathbf{s})-E(\mathbf{s}^\prime)}{T}} \right)
\label{eqn:MH_Q_cancel}
\end{equation}

Of course, this elegant simplification of the Metropolis acceptance criteria is not enough \change{to create a useful quantum algorithm}. Both perturbation theory and numerical analysis have shown that a quantum update as above have some remarkable properties \cite{layden2023}. Proposals that maximize the Hamming distance between $\mathbf{s}$ and $\mathbf{s}^\prime$ (as in the uniform update) while minimising the Energy difference (as in the local update) and are thus likely to be accepted are optimal. This is exactly the criteria that the QeMCMC meets, meaning it has freedom of movement on the lattice at no cost to acceptance probability. One can view this advantage as ``searching'' over a superposition of possible states of similar energy, before collapsing to one at measurement. This becomes incredibly important when in the low temperature regime. As mentioned in Section \ref{sec:MCMC}, Markov-chain suffer from low acceptance rate in the low temperature regime so proposals that can suggest similar energy states are highly likely to be accepted and can significantly improve a Markov-chains convergence.

\section{Framework and figures of merit} \label{sec:fwork_fom}

In this section, we establish the foundational framework for employing CGQeMCMC techniques on the Ising model by outlining the key components and methodologies. Additionally, we provide a comprehensive overview of two pivotal figures of merit: the absolute spectral gap and magnetization estimation. These metrics are introduced to provide background for the numerical experiments detailed in Section \ref{sec:numerics}.

\subsection{Framework}

In CG modeling, the behavior of a complex system is simulated by using a simpler representation of the system. This is immediately realized in Ising models at each MCMC step by isolating a group consisting of $q$ spins - here referred to by the corresponding fraction of the lattice per group $(q/n)$. These spins may not necessarily be close to one another or even interact, however the optimal (in terms of thermalization) choice of spins in a group is likely to depend on the details of the problem. The coarse graining is shown in Algorithm \ref{alg:CG}, where the QeMCMC (Algorithm \ref{alg:QeMCMC}) is run as a subroutine on a small group $s_g$.

\begin{algorithm}
    \caption{Coarse Graining}
    \label{alg:CG}
    \SetAlgoNoEnd
    \SetAlgoNoLine
    \SetNlSty{}{}{}

    $\mathbf{s} \gets$ initial spin configuration\;
    $q \gets$ size of group\;

    $\mathbf{s}^\prime \gets \mathbf{s} $
    
    $index \gets \text{find spins in group}$\;
    $s_g \gets \text{subset of } s \text{ using } index$\;
    \tcp{Do QeMCMC subroutine, Algorithm \ref{alg:QeMCMC}}
    $s_g ^\prime \gets$ QeMCMC ($s_g$)\; 
    \tcp{update $\mathbf{s}^\prime$ with results from $s^\prime_g$ QeMCMC}
    $\mathbf{s}^\prime[index] \gets s_g^\prime$ \;

\end{algorithm}

Of course, there is scope for a multitude of different groupings, sub-groupings and clusterings as is sometimes seen in classical literature \cite{zhu2015, houdayer2001, wolff1989, swendsen1987, wu2021, murray2007}. These are not exhaustively discussed here, where we simply give the framework of coarse graining the Ising model lattice along with some illustrative examples that demonstrate the potential of the approach. Discussion of the choice of spins to include in the active group, as well as the parameters chosen as input to the QeMCMC, including the choice of the ``reduced'' Hamiltonian, are left to Section \ref{sec:CGQeMCMC}. It should be made clear that any Quantum enhancement found can most likely be combined with traditional MCMC methods in hybrid schemes that are likely to outpace ``plain'' MCMC, regardless of proposal.

\subsection{Spectral Gap}

As mentioned in Section \ref{sec:therm}, estimating the thermalization time $(\tau_{\varepsilon})$ of an algorithm is not simple, as one cannot necessarily tell when the Markov chain is in the stationary state. There is also a dependence on the starting state of the algorithm meaning between successive MCMC runs thermalization can differ in steps by orders of magnitude. The thermalization time is however bounded analytically:

\begin{equation}
\left(\delta^{-1}-1\right) \ln \left(\frac{1}{2 \varepsilon}\right) \leq \tau_{\varepsilon} \leq \delta^{-1} \ln \left(\frac{1}{\varepsilon \min _s \mu(s)}\right) \ ,
\label{eqn:spec_therm}
\end{equation}

where $\varepsilon$ is the error in total variational distance used to quantify the bounds of $(\tau_{\varepsilon})$ \cite{levin2017}. The Spectral gap $(\delta)$ describes the slowest facet of a chains convergence and is found through the transition probability matrix, $P$. Given the eigenvalues $\{\lambda\}$ of $P$, $\delta$ is calculated: 

\begin{equation}
\delta=1-\max _{\lambda \neq 1}|\lambda|
\label{eqn:spec_gap}
\end{equation}

The Spectral gap is inversely proportional to the upper bound of the thermalization time (Equation \ref{eqn:spec_therm}) and thus provides an unambiguous quantification of a Markov chains convergence. Numerical calculation of $\delta$ is very inefficient however, requiring numerical integration (often by random sampling, see Appendix \ref{app:imp}) to create the $(2^n \times 2^n)$ $P$ matrix before Eigen-decomposition. Due to this, calculation of the spectral gap is exponentially more demanding than an actual MCMC algorithm.

The spectral gap also ignores the use of a particular chain in more sophisticated schemes mentioned in Section \ref{sec:therm} that rely on running multiple chains concurrently. Thus, one must be reluctant to overstate the results of spectral gap analysis.

\subsection{Convergence analysis}

To analyze the convergence of CGQeMCMC in test cases that is currently technologically inaccessible to QeMCMC, we run Markov Chains and estimate the average magnetization and energy of the Boltzmann distribution. For low temperatures, this essentially becomes an optimization problem as only a few states contribute. 

As mentioned in Section \ref{sec:preliminary}, calculating the average magnetization $\langle m\rangle_\mu$ is dependent on first calculating the partition function. This makes direct calculation infeasible for even relatively simple systems. The alternative is to perform the following estimation where $m(s)$ is calculated for each sampled state by Equation \ref{eqn:m}:

\begin{equation}
    \langle m\rangle_\mu \approx \frac{1}{N} \sum_{\mathbf{s} \in \mathcal{S}} m(\mathbf{s})
    \label{eqn:m_approx}
\end{equation}

Assuming thermalization of the Markov chain, this is accurate with high probability even when $N$ is very small in comparison to the entire state space \cite{ambegaokar2010, levin2017}. One can use estimation of this observable as a figure of merit by asking the question: ``How well can our Markov chain estimate the true magnetization?''. Of course this requires knowledge of the stationary distribution. Here the stationary distribution of exemplar 16, 25 and 36 spin instances are found by brute force, meaning that the update proposals can be evaluated against the true magnetization as benchmark.

Specifically, this paper will look at cumulative magnetization as sampling from Markov Chains generally requires thinning (to take non-correlated samples) and burning (to ensure thermalization) which incurs hyper-parameters complicating the analysis.

Crucially, in certain cases, magnetization estimation can be a flawed figure of merit. As briefly discussed in Section \ref{sec:therm}, a chain may appear to converge to the stationary distribution, while it is actually stuck in a local minimum. If the magnetization contribution of this local minimum is similar to that of the actual magnetization, then this becomes a flawed figure of merit. Thus, alongside the magnetization calculation we analyse the convergence of average energy between multiple chains which can be compared with the exact average energy for the stationary distribution. If local minima impede convergence, this measure will display an obvious bias.

\section{Coarse graining strategies} \label{sec:CGQeMCMC}

In this section, we give three different coarse graining strategies: (a) a ``naive'' approach where one chooses a random subset of spins and ignores the rest, (b) an improved local approach where one chooses a random subset of spins while treating the remaining spins as an external field and (c) an approach that considers multiple subsets, treating each of them separately as in the earlier approaches. \change{Each of these coarse graining strategies is - to our knowledge - novel and of increasing complexity following our scientific exploration.}

Before proceeding to the proposed strategies, let us highlight two properties of the QeMCMC method that were key to the advantage demonstrated in \cite{layden2023}. The first is that the Hamiltonian used (time evolution of which obtains the proposal state) includes the problem's Hamiltonian, and with suitable choice of parameters the proposal step does not differ significantly in energy -- thus ensuring proposed steps that are accepted even when being close to thermalization. The second is that, unlike classical local strategies, QeMCMC can lead to proposal step that differs (significantly) in Hamming distance from the previous step, ``spanning/searching'' the full space and escaping from local minima.

\subsection{Naive CG - local groups}\label{sec:naive_groups}

The naive choice of coarse graining is to pick the \change{indices} of spins that are neighbors forming \change{the set, $g_{l} = \{i_r, i_{r+1} ... i_{r+q}\}$ where $i_r$ is the index of a randomly selected spin $s_r$ that begins our ``local group''.}

\change{The ``locality'' of the spins refer to their indices, which in some cases (for example the 1-d Ising model) also carry physical information, while in other cases (random Ising model) do not. We still use the term ``local'' to distinguish this coarse graining strategy from others, but this should not be misunderstood as ``physically'' local. Future optimal coarse graining strategies will likely not use local groupings, but we limit our analysis in this paper to local groupings for simplicity.}

Once isolated, the group is updated according to the QeMCMC routine outlined in Section \ref{sec:QeMCMC} with $\tilde{J}$ and $\tilde{h}$ the parameters defining the ``partial Hamiltonian'' of the newly formed group. \change{This new Hamiltonian represents only the selected spins, while essentially ignoring the spins not selected:}

\begin{equation}
\tilde{J} = J_{i \in g_l,j \in g_l} \ , \  \tilde{h}  = h_{i \in g_l}
\end{equation}
Ignoring the surrounding spins is an approximation that understandably has limitations and specifically raises two main issues. Firstly, the Hamming distance of proposal steps is fundamentally restricted to the size of the group. While this is better than classical local move approaches, it still puts a bound on the jumps that the proposal steps can make, and is more likely to get stuck in local minima (as with the classical local proposal) and partially cancels one of the key properties that lead to the success of the QeMCMC - the large Hamming distance update proposals.

The second issue is that the new problem Hamiltonian $H_{prob}^{\prime}$ is missing contributions from the ignored spins making the evolution fundamentally incomplete. This partly cancels the other advantage of QeMCMC highlighted earlier, that it includes the problem's Hamiltonian. If one were to consider a 1D Ising model where only neighboring spins interact, this choice of CG is likely to perform adequately, as the boundary interaction terms of the group is constant (2) with cluster size, meaning the impact of missing contributions from ignored spins scales favorably with group size. The same is not true for general Ising models with higher connectivity.

To sum-up, this CG shares some of the advantages of QeMCMC: can ``tunnel'' through local minima (but with limitations on the width due to the bound on the Hamming distance that the proposal step can make); has some information of the problem Hamiltonian (but misses important information for general Ising Hamiltonians).

\subsection{Improved local groups}\label{sec:imp_group}

\textit{Motivation:} As we have seen above, the local groups CG becomes increasingly inaccurate when considering Hamiltonians with high connectivity, since the contributions from the ``ignored'' spins become more relevant. An important observation, however, is that the non-selected spins are constant with respect to the selected ones, and we could treat their interaction energy as a field.

 Explicitly, the interaction matrix is the same as in the local grouping, $\tilde{J} = J_{i \in g_l,j \in g_l}$, but the non-selected spins are treated as environment, adding them to the field of the $j^{th}$ selected spin as:

\begin{equation}
\change{
\tilde{h}_j  = h_{j} + \sum_{i \notin g_l}^n J_{ji} s_i
}
\end{equation}

\change{Where each $s_i$ (value of spins not in the group) is now a \textit{constant} parameter. The 2\textsuperscript{nd} (field) term of the cost function now becomes:}

\begin{equation}
    \change{
    \sum_{j \in g_l} \tilde{h}_j s_j = \sum_{j \in g_l}  s_j  (h_{j} + \sum_{i \notin g_l} J_{ji} s_i)}
\end{equation}

In this way, the accuracy of the Hamiltonian is increased from the local group method without the addition of any quantum gates. The effective Hamiltonian $H_{prob}^\prime$ now fully represents the \textit{relative} energy of any two states of the group. As will become clear in Section \ref{sec:numerics}, improved local grouping refines the naive local groups, however there is still tail-off for small $q/n$.

Overall, this \textit{Improved} local group method can tunnel through local minima with the same Hamming distance limitations as the local grouping, but now contains a much more accurate representation of the problem Hamiltonian.

\subsection{Multiple groups}\label{sec:mult_group}

\textit{Motivation:} To mitigate the issue of low Hamming distance proposals, one can evaluate multiple local groups in a single step meaning the one-shot QeMCMC becomes a succession of single shots of multiple local groups.

\begin{algorithm}
    \SetAlgoNoEnd
    \SetAlgoNoLine
    \SetNlSty{}{}{}

    \tcp{Initialization}\label{alg:CGQeMCMC}
    $\mathbf{s} \gets$ initial spin configuration\;
    $n_g \gets$ number of groups\;
    $l_g \gets$ size of group\;

    $\mathbf{s}^\prime \gets \mathbf{s} $\;
    \For{$i \gets 0$ \KwTo $n_g$}{
        $index \gets \text{spins in } i^{th} \text{ group}$\;
        $s_g \gets \text{subset of } \mathbf{s} \text{ using } index$\;
        \tcp{Do QeMCMC subroutine, Algorithm \ref{alg:QeMCMC}}
        $s_g ^\prime \gets$ QeMCMC ($s_g$)\; 
        \tcp{update $\mathbf{s}^\prime$ with results from $i^{th}$ group's QeMCMC}
        $\mathbf{s}^\prime[index] \gets s_g^\prime$ \; 
    }

    \caption{CGQeMCMC subroutine}
    
\end{algorithm}

Essentially, instead of taking just one local group of spins, we evaluate multiple (disjoint) groups, each individually. While there are many such strategies, we choose a simple one that reduces asymptotically the size of QPU required: we use $\sqrt n$ groups of size $\sqrt n$. This method can be easily combined with the improved local group method resulting in an iterative process that loops through the entire lattice where each quantum computation inherits the measurement results of the last. 

Explicitly, we first split the spins into our $\sqrt n$ local groups before evaluating the first group individually using the Improved local group method. The measurement results of this group then update the state of the system, impacting the environment of the next group which is subsequently evaluated using the Improved local method. In this way, our algorithm iterates through the groups, where at each step the measurement results update the system for subsequent groups' update. The MH acceptance is only computed after the entire state has been updated by each of the  $\sqrt n$ groups. For the remainder of the paper, this method will be labeled ``multiple sampling'' and will harness the Improved local group strategy.

With this CG we are able to maintain both advantages of QeMCMC described at the start of the section: can jump to arbitrary Hamming distances; use the full problem Hamiltonian for these jumps.

\section{Numerical Results}\label{sec:numerics}

In this section we analyse the results of our numerical experiments of Ising model CGeMCMC, while further details of the implementation can be found in Appendix \ref{app:imp}.

\subsection{Spectral gap}\label{sec:numerics_spec}

\begin{figure}[ht]
	\centering \includegraphics[width=0.4\textwidth]{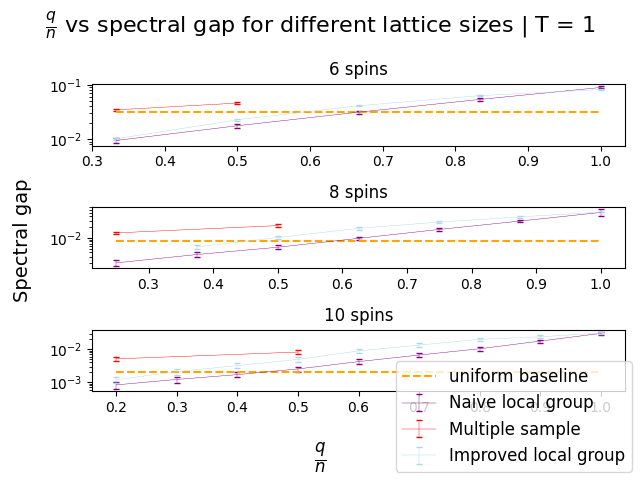}	
	\caption{Spectral gap with $q/n$ for $ n = {6,8,10}$, with classical uniform benchmark in orange. Naive local group (purple) is seen to perform badly as $q$ is decreased, while using an improved local group (light blue) performs significantly better. Multiple sampling with Improved local groups (red) is seen to perform considerably better than both, however the existence of few very factors of small numbers limits proper analysis. Note that the logarithmic scale on the y-axis varies between subplots.}
	\label{fig:spec_gap_interp}%
\end{figure}

\begin{figure}[ht]
	\centering 
	\includegraphics[width=0.4\textwidth]{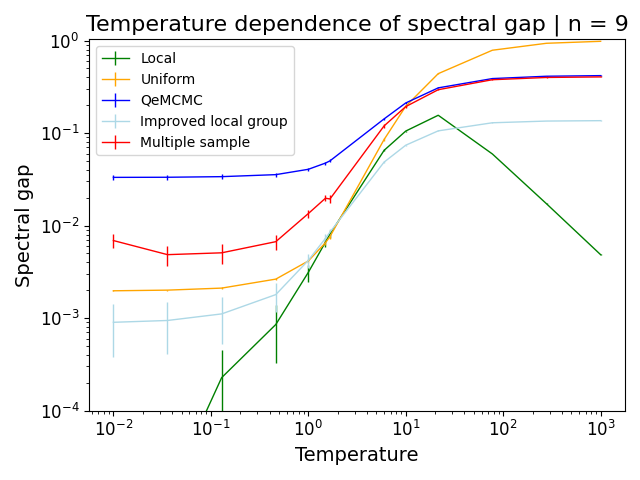}	
	\caption{Spectral gap varying with temperature for $100$ randomly initialized $9$ spin fully-connected Ising model. The classical uniform (orange) and local (green) strategies are clearly improved at low temperatures by the QeMCMC strategy (blue) \cite{layden2023}. The red data represents CGQeMCMC with $q = \sqrt{n} = 3 = n_g$ ($3$ groups of $3$ spins). The light blue data is $q = \sqrt{n}$ but $n_g = 1$ ($1$ group of $3$ spins).} 
	\label{fig:9_T_range}%
\end{figure}

To quantify the CGQeMCMC, the effect of reducing $q$ on the convergence of the resulting Markov chain must be understood, with the particular aim of understanding the different coarse grained proposals at $q = \sqrt{n}$. This is done by evaluating the spectral gap of the CG methods for randomly initialized fully-connected Ising models in the range $4-10$ spins. Figure \ref{fig:spec_gap_interp} shows the decrease of spectral gap with $q/n$ for each of the CG methods and $ n = \{6,8,10\}$. Clearly the decay is worse for techniques that do not multiple sample, suggesting that the locality of low $q$ restricts the proposal. This effect is dependent on system size, however it is difficult to quantify using only the small systems at our disposal. There is also a temperature dependence, as the performance of small $q$ is inextricably linked with the performance of the classical local proposal. For analysis of temperature dependence for a 9 spin instance see Figure \ref{fig:9_T_range}.

In order to assess how the multiple sample technique fares, one must be careful. A silent parameter here is the remainder of integer division in $q/n$. If we have $n = 10$ and $q = 6$, the remainder is 4, meaning an optimal multiple sample strategy with $q = 6$ also has another 4 qubit quantum computation. As this is an effect that is negligible at large n, we avoid this complication by finding the two closest integer values to $q = \sqrt{n}$ and interpolating to estimate the spectral gap value. The interpolation is linear, with errors propagated from the error in spectral gap of the two neighbouring points. The linear assumption is likely to underestimate the true value. Of course, for 4 and 9 spins interpolation is not required.

In analogy with \cite{layden2023} Figure 2a, the different update proposal strategies are compared for a representative set of $100$ randomly initialized $9$ spin fully-connected Ising model instances in Figure \ref{fig:9_T_range}. The QeMCMC (blue) is far superior to any classical strategy (orange and green) in the low temperature regime. The CGQeMCMC with $3$ groups of $3$ qubits (red) is also superior to the classical strategy, however the speedup is clearly dampened. If one evaluates a single $3$ spin group on the lattice each quantum proposal (transparent blue) the quantum speedup is lost for most temperature ranges for this system size.

\begin{figure}[ht]
\centering 
	\includegraphics[width=0.4\textwidth]{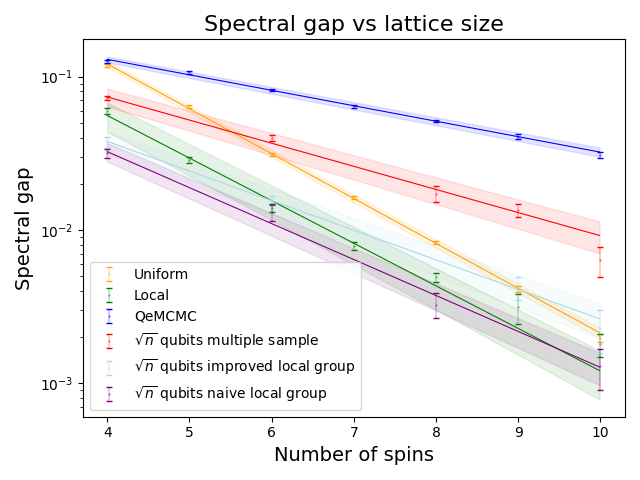}	
	\caption{Spectral gap varying with problem size for $500$ randomly initialized fully-connected Ising models of each integer number of spins. Only $50$ systems of $9$ and $10$ spins for CG techniques were evaluated due to computational limitations. Each data set has least squares exponential fits with fitting function Equation \ref{eqn:fit}. As the system size increases, the spectral gap of classical uniform (orange) and local (green) strategies clearly decrease faster than those of the QeMCMC (blue) and CGQeMCMC (red, light blue) proposals. For a given proposal, the value of $k$ determined by best fit is as follows: Uniform: $0.97(1)$, Local $0.92(4)$, QeMCMC: $0.335(8)$, local group:  $0.77(2)$, Improved local group: $0.64(2)$, multiple sample: $0.50(3)$. $k$ values for $T = 0.1$ and $T = 10$ are listed in appendix \ref{tab:all_T_res} alongside the appropriate figures.}
	\label{fig:T1}%
\end{figure}

There is evident promise of quantum speedup using merely $\sqrt{n}$ qubits, however speedup at a fixed system size is less important. 
The scaling behaviour of the proposal step is necessarily analysed at $T = 1$ in Figure \ref{fig:T1}. Equation \ref{eqn:fit} is the exponential fitting function used in the figures, where a small $k$ represents good scaling of a proposal strategy. The data points for larger systems are less reliable 
due to a much restricted sample size $(50)$ for the average spectral gap. 
This leads to added uncertainty in the fits, and increased reliance on the convergence analysis in Section \ref{sec:numerics_conv}.

\begin{equation}
\delta = a \times 2^{-kn}
    \label{eqn:fit}
\end{equation}

In Figure \ref{fig:T1}, the classical uniform and local proposals perform poorly, with $k = 0.97(1) \ , \ 0.92(4)$ respectively. The coarse grained quantum proposals perform considerably better, in particular the multiple sample proposal with $k = 0.50(3)$. The average quantum enhancement factor ($k_{QEF}$) is the ratio of $k$ for the quantum algorithm and the $k$ from the best classical proposal:

\begin{equation}
k_{QEF} = \frac{k_{classical}}{k_{quantum}}
\label{eqn:q_e_f}
\end{equation}

For the multiple sample proposal strategy, we see a quantum enhancement factor of $ k_{QEF} = 1.84$ against the best classical approach - the local proposal. In this data, the QeMCMC gains an enhancement of $2.7$, a slightly reduced value than reported in \cite{layden2023} for the same temperature. Figure \ref{fig:T1} depends largely on the temperature while Figure \ref{fig:9_T_range} depends on the problem size. Appendix \ref{app:extra_data} provides multiple figures that explores these relations. The corresponding quantum speedups in $k$ for $T = 0.1$ and $T = 10$ are $2.17$ and $2.6$ respectively. Of course, the problem instance type (1D, 2D, fully connected, sparse etc.) influences the outcome of these experiments, however these topics are beyond the scope of this work.

Figure \ref{fig:T1} may wrongly suggest that the uniform proposal is a useful step. In practice, the local proposal is much more useful in the classical schemes such as parallel tempering mentioned earlier. Any quantum proposal step, particularly when the Hamming distance of steps has been reduced by CG, will be useful in much the same way that the local step is in practice.

\subsection{Convergence analysis}\label{sec:numerics_conv}

Markov chains using local, uniform and CGQeMCMC are used to evaluate the average energy and magnetization of the stationary distribution of exemplar fully-connected Ising models. Systems of size $16$, $25$ and $36$ (see Appendix \ref{app:extra_data} for $15$ and $25$ spin results) are the focus so that $\sqrt{n}$ qubits can be easily compared with classical methods. The stationary distribution of the $36$ spin instance at $T = 0.1$ is the primary focus of this analysis as it is the largest system studied and has low temperature, which as seen in Figure \ref{fig:9_T_range} is where we expect CGQeMCMC to perform well. The problem is approaching that of an optimization problem (the limit reached at zero temperature), as most of the probability distribution lies in the ground state of the system, making it particularly difficult. See Appendix \ref{app:extra_data} for temperature and system size analysis.

\begin{figure*}
	\centering 
	\includegraphics[width=1\textwidth]{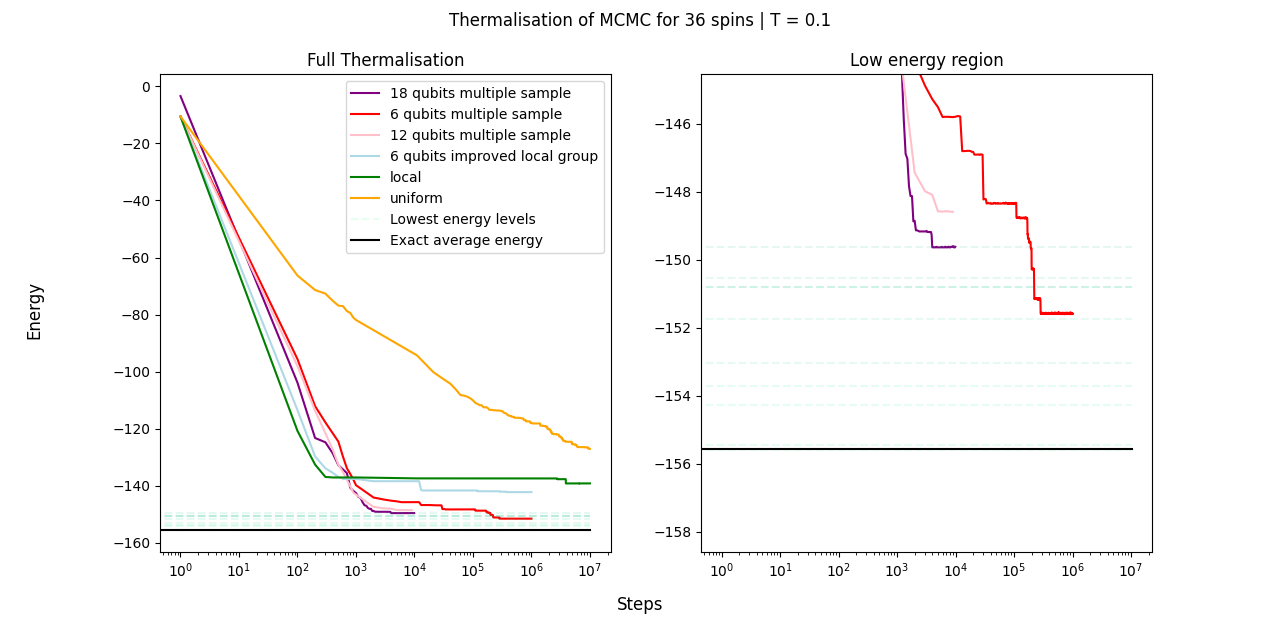}	
	\caption{ Average energy of $20$ chains at $100$ step intervals for uniform (orange), local(green), Improved local group without multiple sampling (light blue) and multiple sampling (red) with $q = \sqrt{n} = 6 = n_g$ ($6$ groups of $6$ spins) for $T = 0.1$. Multiple sampling with $q > \sqrt{n}$ is also simulated, with 12 qubits (pink) and 18 qubits (purple). Right figure is close-up of the low energy region. The more local a proposal, the larger the offset from the true average energy (black line). This is greatly reduced by CGQeMCMC with multi-sampling and larger groups - methods that decrease the locality of proposals. The uniform proposal suffers from extremely slow convergence but does not get stuck in local minima. }
	\label{fig:E_36}%

\end{figure*}

Figure \ref{fig:E_36} shows that all proposed methods are unable to converge within simulable timescales. The CGQeMCMC appears most effective, even though 
the problem of getting stuck in local minima persists 
since the average energy over $20$ Markov chains is biased by a few that have not yet ``found'' the minimum state. One can crudely evaluate convergence in this relatively low temperature regime by calculating how many chains have discovered the minimum state. In this example, where we consider $20$ chains for each proposal, $5$ local chains, $0$ uniform chains and $11$ CGQeMCMC chains (6 qubit multiple sampling) have found the minima. 
The classical methods had an extra order of magnitude steps (due to computational limitations) to search the state space, and thus it is evident that quantum proposals are significantly more efficient. When one considers a Markov chain as part of a larger algorithm, with multiple parallel chains, 
this improvement in the quality of convergence becomes more substantial, 
even in the cases that it still fails to 
converge perfectly.

It should also be noted that increasing the number of qubits used (smaller coarse graining) has a positive effect on convergence. Simulations of 12 qubits (pink) and 18 qubits (purple) show increasing promise, despite being 
simulated for far fewer steps. This highlights the flexible nature of this coarse graining algorithm, that can take as input the size of the hardware that one has and accordingly achieve the best possible performance.

\begin{figure}[ht]
	\centering 
	\includegraphics[width=0.4\textwidth]{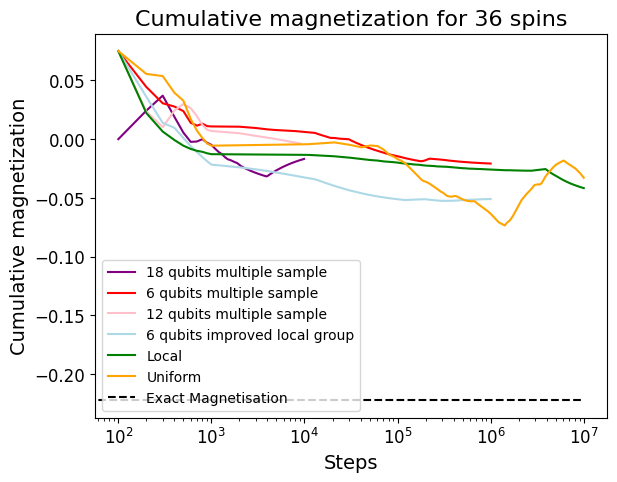}	
	\caption{Average cumulative magnetization of $20$ chains of each proposal method where the colors are the same as in Figure \ref{fig:E_36}. If the exact magnetization were not known, one may erroneously believe that some methods may have converged, when in reality none have.}
	\label{fig:M_36}%
\end{figure}

The magnetization plots confirm lack of convergence of all proposed methods when compared with the brute force exact magnetization. It is important to note, that one cannot infer from the magnetization plots the convergence. Specifically, while a failure to converge in the magnetization plot also implies failure of convergence in general, the converse does not hold, since it is possible to have instances that appear to get the correct magnetization but fail to converge to the correct probability distribution. 
This becomes more important 
for ``real life'' examples where brute-force numerics are not possible.

It is clear that even when employing only $6$ simulated qubits, a quantum enhanced \change{exploration} of the state space is more effective than both classical methods in the low temperature regime. One should note that when the temperature is raised to $T = 1$, this quantum \change{enhancement} is diminished. See Appendix \ref{app:extra_data} for the complete set of plots for $T = \{0.1, 1\}$ and $n = \{16, 25, 36\}$. 

Although the results reported in this section point towards to a scaling advantage, this cannot be explicitly proven by analysis of only a select few Ising model instances. A more thorough analysis with multiple different models would be required.

\section{Conclusions}\label{sec:conc}

In this paper starting with the QeMCMC approach of \cite{layden2023}, we identified two important elements that lead to its desirable performance, namely the ability to suggest proposal steps with big Hamming distance from the previous configuration (due to generating superposition of all configurations), while returning a new configuration that is close in energy with the previous one (due to the use of the problem Hamiltonian and the suitable annealing-style brief evolution). \change{At the same time, we identified a limitation of QeMCMC regarding the requirement of a QPU with as many qubits as the size of the} Ising system considered. 

In order to enable the use of small QPUs we introduced the concept of coarse graining, meaning that we considered a subset of the overall system in order to find the next suggested configuration. To make our proposal competitive with QeMCMC and classical solutions, we aimed to find a coarse graining that retains (largely) the two advantages of QeMCMC. Our final ``multiple groups'' coarse graining achieves this: can give rise to new configurations that differ as much in Hamming distance as one wishes; for each ``local'' subpart considered we used an effective reduced Hamiltonian that treats the coarse grained spins as an external field, achieving a sufficiently good approximation.

We then tested our proposal(s) using two figures of merit: The absolute spectral gap and magnetization estimation. Spectral gap analysis of simulated quantum proposals in Section \ref{sec:numerics_spec} was confined to systems comprising 10 spins or fewer. Nevertheless, the observed scalability of CGQeMCMC proposals relative to system size shows promising trends. Using only $\sqrt{n}$ qubits, Quantum-enhanced Markov Chain thermalization is predicted to outperform classical techniques for larger systems, giving a quantum enhancement factor $k_{QEF}$ (see Section \ref{sec:numerics_spec} and Equation \ref{eqn:q_e_f} specifically) between $1.84$ and $2.6$. 

The Markov Chain Monte Carlo algorithm was then executed in order to estimate the magnetization for larger systems up to 36 spins, with coarse grained quantum proposals again using only $\sqrt{n}$ simulated qubits. Both classical proposals analysed here (local and uniform) struggle to reach thermalization in these larger systems, falling into local minima or inefficiently exploring the large state-space. The coarse grained quantum proposals improve convergence overall for these larger systems, however they they also occasionally suffer from local minima.

There are many directions for future investigations related to the CGQeMCMC. The choice of $q = \sqrt{n}$ qubits in this paper was successful, however the optimal relation between $q$ and fastest thermalization or indeed auto-correlation \change{is both problem and hardware dependent. Future work should use the CGQeMCMC to exploit the connectivity of hardware. For example, if one were to consider an Ising model with 3D connectivity, they could partition the problem into $n$ planar lattices that are suitable for the 2D connectivity common to many QPU. In this way, the CGQeMCMC can be a powerful tool in relating highly connected problems to QPUs with limited connectivity.}

The effect of noise has not been investigated. The QeMCMC method is by construction noise-resilient, and our CG approach uses smaller QPUs minimizing further these effects. Concrete analysis for the level and types of noise \change{under which we maintain the quantum scaling advantage is an important open question.} Here it is also worth pointing out that due to the
one-shot nature of the algorithm, standard error mitigation methods \change{are not applicable} meaning it would require a bespoke scheme that is outwith current literature \cite{li2017, temme2017}.

\change{Another future direction is quantification of the sampling error induced by coarse graining. A Markov chain produced by the CGQeMCMC provably samples the Boltzmann distribution as it satisfies detailed balance and is aperiodic and irreducible. The number of steps required to produce enough samples to estimate some observable within a given error is dependent on the auto-correlation time of the chain. Speeding up thermalization does not necessarily result in quicker auto-correlation, and thus this must be studied in future work to properly understand the cost of coarse graining compared to the QeMCMC.}

An important future step is to apply coarse graining to problems other than the Ising model. Application of variationally prepared Hamiltonian evolution is also an interesting avenue of research, as is coarse graining quantum algorithms that have already been derived from the QeMCMC \cite{lockwood2024, nakano2023}. Other interesting next steps are to apply the CGQeMCMC to quantumly enhance classical methods such as simulated annealing, parallel tempering and population annealing.

\section*{Acknowledgements}

PW acknowledges support by EPSRC grants EP/T001062/1, EP/X026167/1
and EP/T026715/1, STFC grant ST/W006537/1 and Edinburgh-Rice Strategic Collaboration Awards and SF acknowledges support by EPSRC DTP studentship grant EP/W524311/1.


%

\appendix

\section{Implementation details}\label{app:imp}

The numerical quantum simulations were carried out using python 3.11, running qulacs quantum circuit simulator \cite{suzuki2021} and can be found open source on Github \cite{Ferguson_2024}. The base of the code was built upon a pre-existing library, however it was drastically modified to fit the requirements of this paper \cite{quMCMC_git}. 

To keep consistent with the original work, parameters have been kept constant as much as possible \cite{layden2023}. For example, $\gamma$ is varied in the range $(0.25, 0.6)$ and the integer time parameter is varied between $(2,20)$ for all of the following experiments. One key difference is the method of calculating the spectral gap for quantum proposal methods. The authors of \cite{layden2023} employed exact numerical integration by discretizing the gamma parameter. Here however, we employ Monte Carlo integration by uniformly sampling the $t$ and $\gamma$ regions. Explicitly, this means that for a given row of the $Q$ matrix, the measurement statistics are estimated by averaging over $m$ (chosen to be 30 by experimentation) randomly generated $t$ and $\gamma$ values. When one is doing plain QeMCMC or CGQeMCMC without multiple groups, the $i^{th}$ row of $Q$ can be found by doing the following $m$ times: First, randomly generate $t$ and $\gamma$ values then use as quantum input the $i^{th}$ state, run the appropriate quantum circuit and output the statevector. Averaging over the $m$ statevectors gives a good approximation of the likely transitions between states. 

For CGQeMCMC however, the situation is more difficult. As the output of one circuit affects the gates of the next, then the entire $Q$ matrix must be found by brute force. First, randomly select a starting state, run a CGQeMCMC step where $t$ and $\gamma$ are randomly chosen and then add $1/n_s$ to the $(i,j)^{th}$ element of $Q$. Repeating this many times leads to an accurate approximation of $Q$. To justify how many times is required, this process was repeated for over multiple different sample numbers with $n_s = (2^n)^2$ deemed fit.

\section{Extra data}\label{app:extra_data}

Here, extra data is included without detailed analysis for the interested reader.

\subsection{Hamming distance and Energy difference of proposals}

\begin{figure*}
	\centering 
 \includegraphics[width=0.9\textwidth]{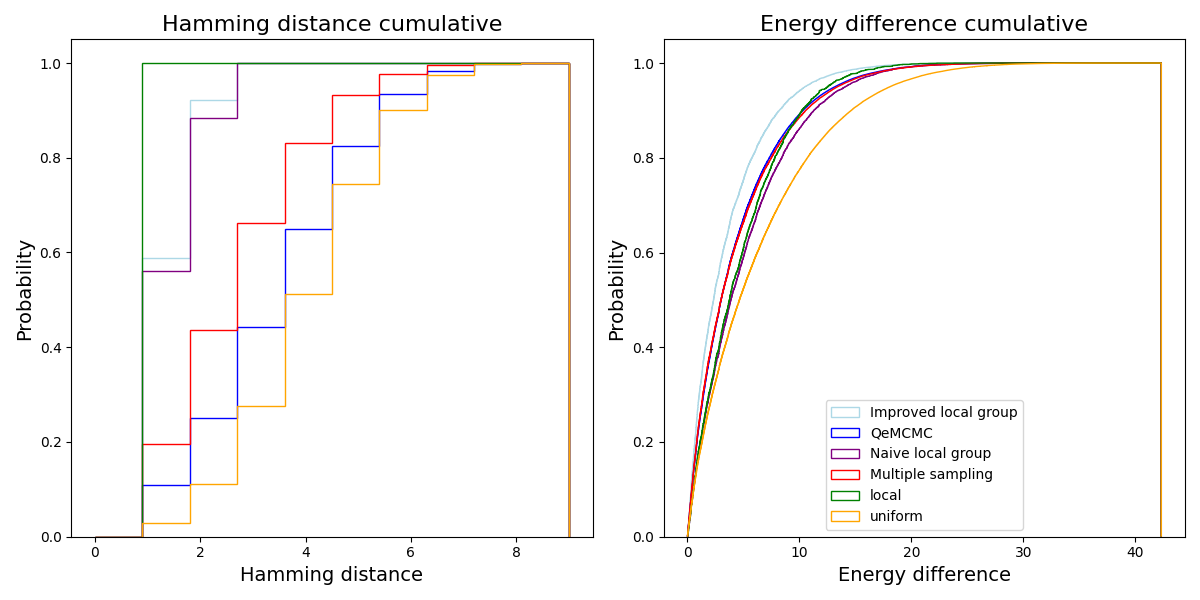}	
	\caption{ Cumulative distributions for an example $9$ spin problem instance, where the Hamming distance (left) and energy difference (right) of each proposal method are compared. For all CG methods, $q = 3$.}
	\label{fig:H_E}%
\end{figure*}

In this section, the Hamming distance and energy difference of proposed steps are compared through Figure \ref{fig:H_E}. The classical local step will only ever take steps of Hamming distance $=1$, meaning it is entirely local but has a high acceptance rate as the energy difference of proposed steps is low. This means that such updates traverse the cost function landscape in a very local way, often getting stuck in local minima for very long periods of time. The uniform proposal has the extreme opposite feature, as it picks a possible step at random, it has a very low acceptance rate meaning slow exploration of the landscape as the energy of a proposed state is unlikely to be close or lower. When a step is accepted however, the hamming distance is very large. Both of the above points are made clear in Figure \ref{fig:H_E}. 

Any CG method that does not multiple sample will have the locality issue of occasionally getting stuck in minima. If one does multiple sample, it becomes clear that the Hamming distance is slightly reduced compared to the uniform or QeMCMC method. The energy difference of proposals still outperforms classical methods, almost matching the QeMCMC. One interesting artefact is that the improved local group (without multiple sampling) greatly reduces the energy difference compared to all methods, while increasing the Hamming distance compared to the classical local method. This means that if one were to choose a problem where the local method performs well (for example in particular temperature regimes) then the improved local method is always likely to sustainably outperform the classical proposals. Scaling of this artifact is left to future work.

\subsection{Temperature and convergence for large systems}
\begin{figure*}
\includegraphics[width=0.3\linewidth]{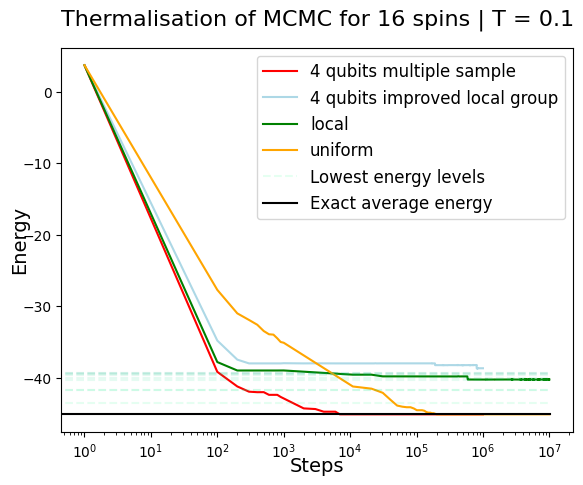}
\includegraphics[width=0.3\linewidth]{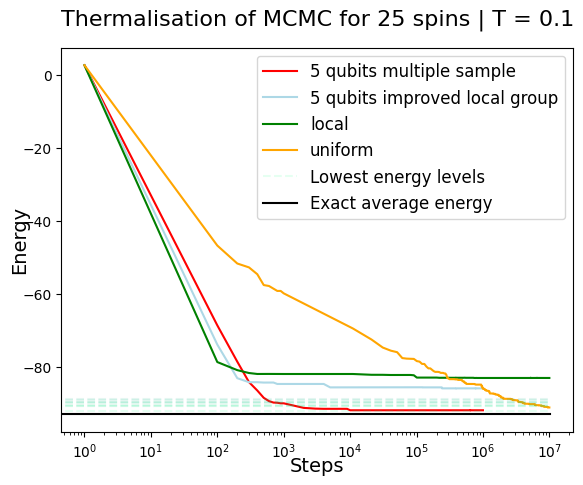}
\includegraphics[width=0.3\linewidth]{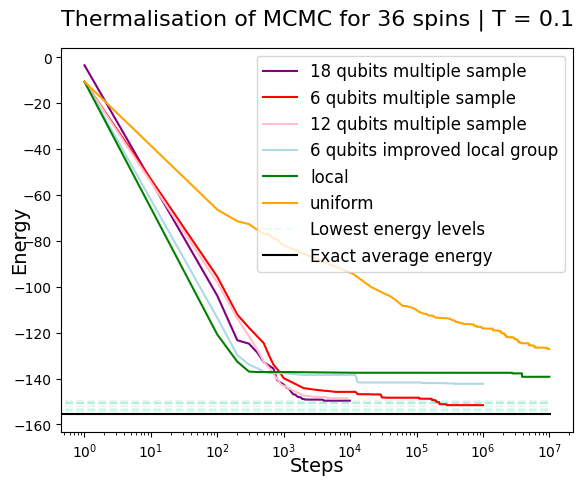}\par 
\includegraphics[width=0.3\linewidth]{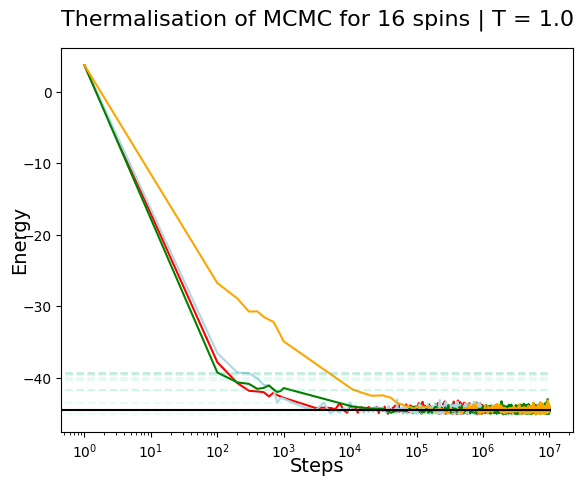}
\includegraphics[width=0.3\linewidth]{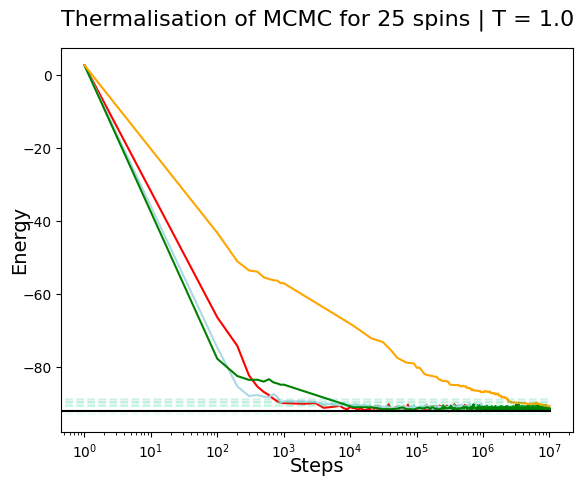}
\includegraphics[width=0.3\linewidth]{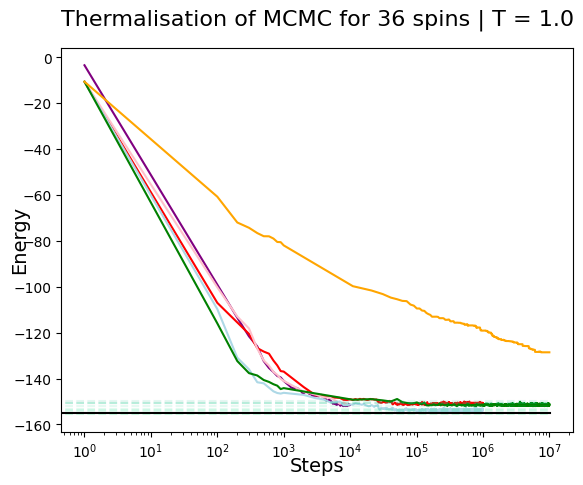}\par 

\caption{Average energy thermalization plot of $20$ Markov Chains at $100$ step intervals for one randomly selected Ising model of size $16$ (left), $25$ (middle) and $36$ (right). $T = 0.1$ (top) is clearly a more difficult problem than $T = 1$ (bottom) where the Markov chains acceptance rate is much higher. The colored dotted lines represent the $10$ lowest energy levels of the system, while the black line represents the exact average Boltzmann energy. 
}
\label{fig:app_therm_all}
\end{figure*}

The relationship between temperature, system size and thermalization is conveyed through analysis of the energy convergence in Figure \ref{fig:app_therm_all}. One can see that the thermalization of the uniform proposal becomes useless for large systems, the local proposal struggles at low temperatures and in every case, both quantum proposals analysed efficiently find a better than classical low energy regime. magnetization plots are not provided as they falsely suggest convergence of non-converged distributions.

\subsection{Temperature and scaling}
\begin{figure*}
	\centering 
	\includegraphics[width=1\textwidth]{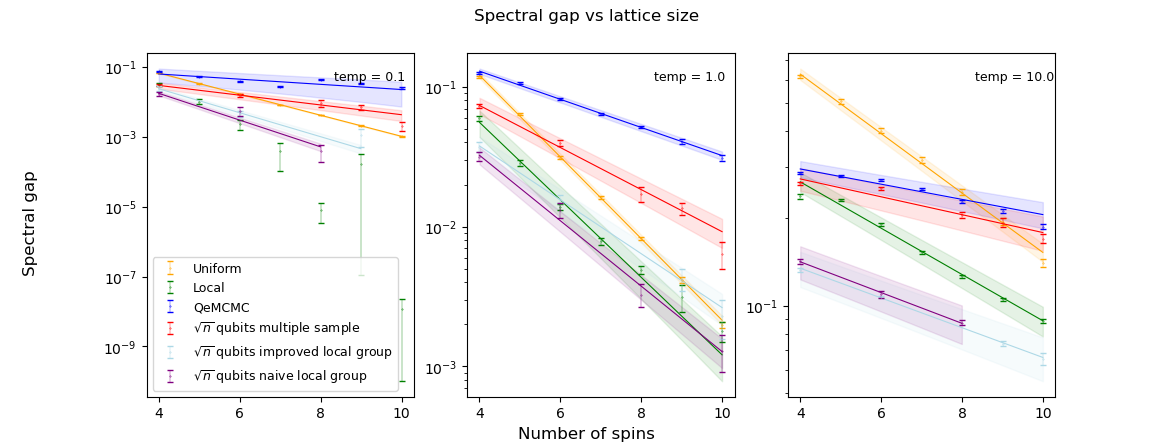}	
	\caption{Scaling of spectral gap with system size for a range of $T$. Fitting results are given in Table \ref{fig:all_T} Classical uniform (orange) and local (green) clearly scale worse than the QeMCMC (blue) as system size increases. Local group (purple) perform variably and the improved local group (light blue) performs better, consistently out-scaling classical methods. Multiple sampling enables (red) CGQeMCMC to almost match QeMCMC in high temperatures, however performance is reduced in low temperatures. } 
\label{fig:all_T}%
\end{figure*}

The temperature dependence of spectral gap scaling is briefly conveyed here through a cross-section of temperature regimes. Figure \ref{fig:all_T} shows the scaling behavior of each proposal method for $T = \{0.1,1\}$ while Table \ref{tab:all_T_res} shows results of each fit according to Equation \ref{eqn:fit}.

\begin{table*}[ht]
    \centering
    \begin{tabular}{|c|c|c|c|}
        \hline
        Temperature & 0.01 & 1 & 10 \\
        \hline
        Uniform & $k = 1.003(2)$ & $k = 0.97(1)$ & $k = 0.341(8)$ \\
        Local & N/A & $k = 0.92(4)$ & $k = 0.26(1)$ \\
        Quantum & $k = 0.25(8)$ & $k = 0.335(8)$ & $k = 0.087(1)$ \\
        Quantum $q = \sqrt{n}$ local group & $k = 1.27(2)$ & $k = 0.77(2)$ & $k = 0.177(5)$ \\
        Quantum $q = \sqrt{n}$ improved local group & $k = 1.1(2)$ & $k = 0.64(2)$ & $k = 0.171(5)$ \\
        Quantum $q = \sqrt{n}$ multiple sample & $k = 0.46(4)$ & $k = 0.50(3)$ & $k = 0.10(1)$ \\
        \hline
    \end{tabular}
    \caption{Table of results for fits in Figure \ref{fig:all_T}}
    \label{tab:all_T_res}
\end{table*}

\end{document}